\documentclass[amsmath,amssymb,lengthcheck,aps,prl] {revtex4}
\usepackage[colorlinks,allcolors=blue]{hyperref}
\usepackage[T1]{fontenc}
\usepackage[utf8]{inputenc}
\usepackage{graphics}
\usepackage{graphicx}
\usepackage{color,soul}
 \usepackage{soul}
\begin{document}
\title{Double roton-minima in bosonic fractional quantum Hall states}
\author{Moumita Indra $^1$ \footnote{Orcid ID: 0000-0002-7900-7947}} 
\email{moumita.indra93@gmail.com / moumitaindra@ee.iitb.ac.in}
\author{Deepak Jain $^{2}$} 
\author{Sandip Mondal$^1$} 
\email{sandip@ee.iitb.ac.in}
\affiliation{$^1$ Indian Institute of Technology Bombay, Mumbai 400076, India}
\affiliation{$^2$ Indian Institute of Technology Delhi, Hauz Khas, New Delhi 110016, India}
\begin{abstract}
We have studied the collective spin-conserving collective excitation spectra in rotating diluted ultra-cold Bose atoms. Double roton-minima have been observed in the fractional quantum Hall (FQH) states for the two filling fractions ($\nu$) of the first series of Jain’s composite fermion sequences. The obtained roton-minima for $\nu = 1/4$ are at the wave-vectors $1.26$ and $2.38$ and the roton-minima for $\nu = 1/6$ have been shifted to $1.08$ and $2.06$. Such shift of roton-minima is attributed due to strong correlation between the particles in bosonic FQH-system. Moreover, the number of roton minima observed depends upon number of attached fluxes as well as the ranges of interaction between the particles.
 
\end{abstract}
\maketitle
Collective excitation of strongly correlated fractional quantum Hall (FQH) fluids is an insightful investigation in a two-dimensional electron system (2DES). Strongly correlated many-body quantum phenomenon, such as the fractional quantum Hall effect (FQHE) \cite{fqhe, fqhe1, fqhe2, QHE} can be explained with the concept of quasi-particles. 
Composite fermions (CFs) \cite{CFs} are the hypothetical quasi-particles introduced by J. K. Jain to explain the fractional plateus \cite{plateau}, obtained in Hall resistivity.
It is assumed that electrons attach even number of flux quanta with themselves from the external magnetic field and form new quasi-particles, named CFs. The majority of the FQH-states observed in experiments may be converted to weakly-interacting integer quantum Hall states with the aid of CF theory. In 1995, Davis \cite{Davis_1995} and others \cite{BEC} became the first to produce the  Bose-Einstein condensates (BEC) by magnetically confined alkali atoms.
Following that, Cooper and Wilkin \cite{CF_on_BEC} investigated the first FQHE of a charged Bose system using the Jain's CF model. Recent developments in rotating BEC \cite{FQHE_BEC} have provided an intriguing framework for exploring FQHE. Rapidly rotating BEC, trapped in a a two-dimensional (2D) harmonic potential creates a fictitious magnetic field ($B$), perpendicular to the 2D-plane that resembles like 2DES in perpendicular magnetic field. Landau levels (LL) are formed and then there should be a high chance to observe FQHE in fast-rotating BEC of  diluted ultra-cold Bose atoms \cite{CF_on_BEC}. As the bosonic system is applied high rotation, so most of the particles fly out, which makes the atomic density of the system very low. In this situation of extremely low atomic density, bosons may exclusively dwell in the lowest LL (LLL). Since the kinetic energy becomes constant, the problem  will be accurately determined by only the interactions potential. At low energy limit only binary collisions are significant in a diluted cold gas and the s-wave scattering length \cite{s_wave} is the only characteristic used to describe these collisions. In this instance, a suitable expression for the interaction term is difficult to formulate. So, short-ranged interactions like $\delta$-function type potential \cite{Delta_pot} between the atoms can be considered. Jain, Regnault, and other theoretician \cite{QH_spinlessBOSON} studied the interacting bosons using CF theory and considered short-ranged $\delta$-function like interaction between the Bose particles. The correlated FQH-states produced in rotating BEC may be explained by the weakly-interacting CFs of Bose atoms, which are the bound states of Bose atoms and the odd number of flux quanta (or quantized vortices) \cite{CF_on_BEC}. Since the vortices tied to the atoms create Berry phases that partially cancel the Aharonov-Bohm phase produced by applied magnetic field; this bound state (Bose-CF) is governed by fermionic statistics. The Bose-CFs experience a reduced magnetic field
\begin{eqnarray}
B^* = B - p \rho \phi_0 \;;
\label{EffB}
\end{eqnarray}
here, $B$ represents fictitious magnetic field, $\phi_0$ is the magnetic flux- quantum; while the number density of the Bose atoms is  $\rho$, and $p = 1, 3, 5, $ is the odd number of flux quanta attached. This Bose-CF creates a new type of LLs called $\Lambda$-levels in this weaker magnetic field. The LL filling fraction of the Bose atoms ($\nu$) can be related with the filling fraction of Bose-CF $\Lambda$-level (assumed to be an integer $n$) from the above relation (\ref{EffB}) as
\begin{eqnarray}
\nu = \frac{n}{n p + 1}
\label{CF_seq}
\end{eqnarray}
Using this mapping, we can demonstrate that interacting bosons can be treated like fermions \cite{CP_statistics} and that the filling fractions precisely match the Jain's sequence \cite{Jain_CF}. 
If we consider the particles are in LLL i.e. $n = 1$ then from the above relation (\ref{CF_seq}) we can have $\nu = 1/2, 1/4, 1/6$ fractions using odd values flux quanta $p = 1, 3, 5$ respectively.  \\
For a particular filling fraction, different partially spin-polarized states can be found depending upon the values of Zeeman energy ($E_z$) and cyclotron energy ($E_c$) of the system \cite{partial_spin, partial_spin1, partial_spin2}. Landau-level mixing \cite{LL_mix} becomes crucial in case of partially spin-polarized FQH-states. To explain the partially spin-polarized states CF theory is not sufficient then we need to apply other theoretical techniques \cite{polarized,polarization_MI}. Here we are considering fully spin-polarized state in presence of strong magnetic field i.e. Zeeman energy is much higher than the cyclotron energy.

In this work, spin-conserving collective excitation has been explored for the filling fractions $\nu = 1/4, 1/6$ in bosonic FQH-states. We obtained double roton-minima for both the two fractions. For $\nu = 1/4$ two roton-minima are obtained at the wave-vector $k = 1.26$ and $2.38$; whereas for $\nu = 1/6$ those are shifted to lower values of wave-vector $k = 1.08$ and $2.06$. Such shift of roton-minima has been explained by the strong correlation between the particles in rotating diluted ultra-cold Bose atoms.
\section{Diluted ultra-cold rotating Bose atoms confined in a harmonic trap}
We have considered $N$ weakly interacting Bose atoms of mass $m$, rotating with angular velocity $\Omega \hat{z}$ trapped in a 2D isotropic harmonic trap (XY-plane). Thus the Hamiltonian for the system can be written as,
\begin{eqnarray}
H &=&  \sum_i \frac{p_i^2}{2m} + \frac{1}{2} m \omega^2 r_i^2 - \Omega L_z + V \nonumber \\
&=& \sum_i \frac{1}{2m} (\vec{p}_i-\vec{A})^2 + (\omega-\Omega)L_z + V  \;;
\label{H_boson}
\end{eqnarray}
where $\vec{r_i}$, $\vec{p_i}$ is the position, momentum of i-th atom respectively, $L_z$ is the projection of angular momentum along Z-axis, $V$ is the interaction Hamiltonian between the atoms and the trap energy is $\hbar \omega$. Corresponding vector potential so obtained is $\vec{A} = m \omega (-y, x)$. This indicates the equivalence to the Hamiltonian of a particle experiencing an effective magnetic field \cite{Hamiltonian}. Since the system is sufficiently dilute so that all the atoms are confined in the LLL. If we can assume $\omega = \Omega$, then the problem becomes identical with the FQHE in 2DES.
\section{Interacting potential between diluted Bose atoms}
\subsection*{1. Short ranged contact interaction}
At the low energy limit, the effective interactions between the Bose atoms can be denoted by a constant in momentum space, since the atoms scattered by the s-wave \cite{s_wave} only. Consequently, an instantaneous $\delta$-function may be used to represent the interaction in real space.  It gives rise to the Gross-Pitaevski (GP) equation \cite{GP_equn, GP_equn1} for rotating BEC and super-fluid systems. This interaction in 2D system has the form
\begin{eqnarray}
V = g \sum_{i<j} \delta(\vec{r}_i - \vec{r}_j) \;;
\end{eqnarray}
with $g$ as the strength of the interaction \cite{exact_12} and $\vec{r}_i$, $\vec{r}_j$ is the position coordinate of $i$-th and $j$-th particle.\\
\subsubsection*{\textbf{P$\ddot{o}$schl-Teller interaction}}
In quantum Monte Carlo calculations, handling the $\delta$-function potential is particularly challenging and needs a significant amount of computational resources. Regnault and Jolicoeur \cite{QH_spinlessBOSON} used exact diagonalization method to obtain low-lying states of limited number of bosons interacting with $\delta$-function potential. Chang {\it et. al.} \cite{FQHE_BEC} and other group studied the low-energy excitation spectra by Monte Carlo method using $\delta$-function interaction but for very small number of bosons up to 12. The energy spectra of a large number of particles cannot be calculated using the $\delta$-function potential. So, in our study, we considered the P$\ddot{o}$schl-Teller (PT) interaction potential ($V_{PT} $) \cite{Poschl-Teller} to get over this issue. PT-interaction has the form 
\begin{eqnarray}
V_{PT} = U \sum_{i<j} \frac{2\mu}{cosh^2 \;( \mu r_{ij})} \;;
\end{eqnarray}
where, the parameter $U$ represents interaction strength and $r_{ij} = \mid \vec r_i - \vec r_j \mid $ is the separation distance between two particles. Here $\mu$ denotes a parameter having unit of inverse of the magnetic length $l_B$; where $1/\mu$ measures the width of the interaction. In order to explore the $\mu$-dependent nature of excitation a range of $\mu$-values have been taken into consideration. As we raise the $\mu$-value, then the essence of PT-interaction goes well with $\delta$-function. But for smaller values of $\mu$ the nature of PT potential differs from the contact interaction, as the range of interaction increases with decrease of $\mu$ \cite{SDE_MI, Physica_B}. That's why we have considered $\mu$-values between a specific range in our calculation. 
However, the PT interaction helps us to get our estimated result in thermodynamic limit ($N\rightarrow \infty$) i.e. finite but large number of particles.
 \begin{figure}
 \centering
  \includegraphics[width = 8.0cm]{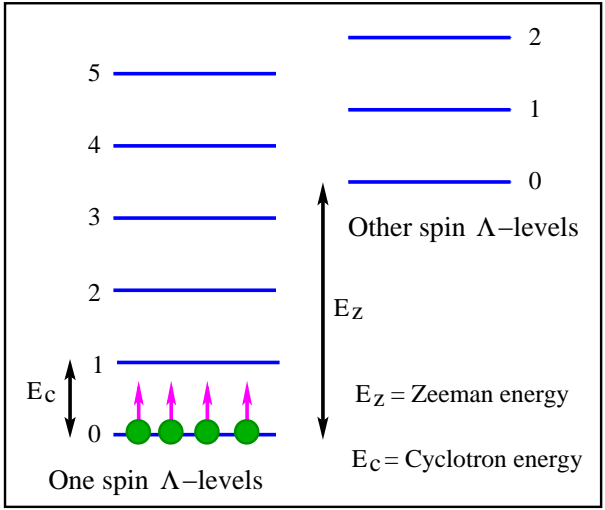}
  \caption{The schematic diagram of ground state for the two-component bosonic filling fraction $\nu=1/2$: The two-component spin states have their own $\Lambda$-levels which are represented by blue straight lines (left panels denotes one-spin states and right panel denotes another spin-states). The number ($0, 1, 2, 3,...$) written beside the lines represents $\Lambda$-level index. Bose-CFs are represented by the green colored solid-dots with pink arrow-lines (Bose atom by solid-dots and magnetic flux attachment by arrow-lines).}
  \label{ground}
 \end{figure}
 \begin{figure}
 \centering
  \includegraphics[width = 8.0cm]{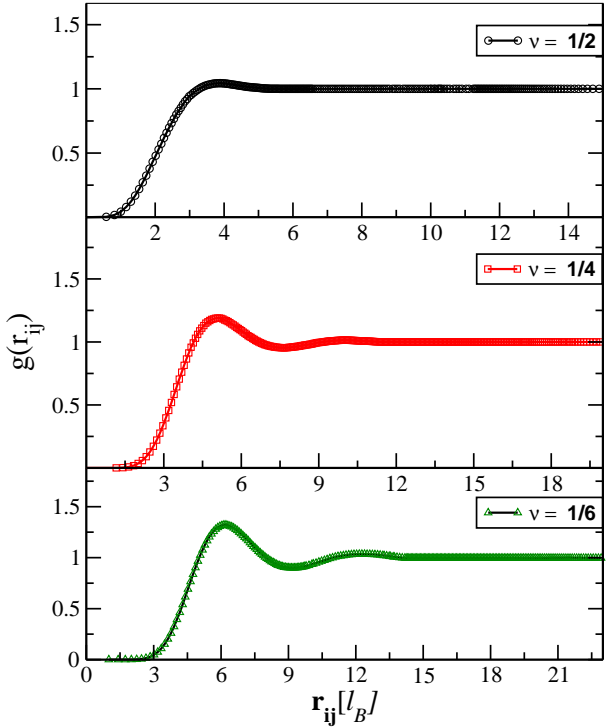}
  \caption{Pair-correlation function $g(r_{ij})$ for the filling fractions $\nu = 1/2, 1/4, 1/6$ are plotted as a function of two particle's separation distance $r_{ij} = \mid \vec r_i - \vec r_j \mid$. For this result sufficiently large number of particles are taken but the nature is independent of the number of particles.}
  \label{PC}
 \end{figure}
 \begin{figure}
 \centering
   \includegraphics[width=8.5cm]{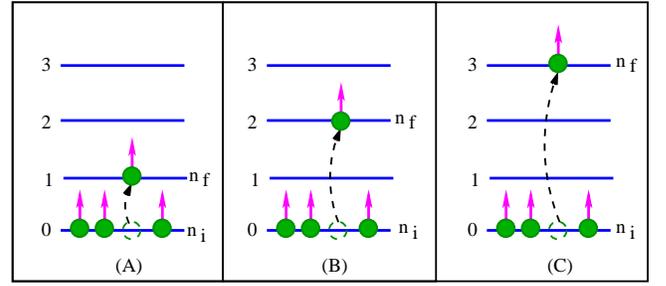}
   \caption{Spin-conserving collective excitation: In each block blue straight lines represent one-component $\Lambda$-levels (CF Landau levels) only. Bosonic FQH-filling fractions of $\nu= 1/2, 1/4,$ and $ 1/6$ translate into one filled $\Lambda$-level with flux attachment $1, 3, 5$ respectively. Therefore, in this figure just one flux quanta is attached, so it represents $1/2$ filling fraction; if there are $3$, $5$ fluxes, it would represent $1/4, 1/6$ fractions respectively. Lowest-order possible excitations for $\nu=1/2$ are presented here, which follows same for other two fractions. These three blocks are the possible spin-conserving excitations that are most likely to occur. The transitions can occur as: (A) 0 $\rightarrow$ 1; (B) 0 $ \rightarrow $ 2; (C) 0 $ \rightarrow $ 3. In each of the three blocks, there are particle-hole pair i.e. CF-exciton.  Missing of a Bose-CF is shown by the dotted-circle and the transition of Bose-CF from $n_i$-th $\Lambda$-level to the $ n_f$-th $\Lambda$-level is shown by dotted arrow-lines.}
   \label{CDE}
 \end{figure}
 \subsection{2. Long range Coulomb potential}
Additionally, in 2DES, FQHE occurs in the presence of a robust Coulomb interaction. We are thrilled to observe collective excitation in a bosonic system for long-range Coulomb interaction ($1/r$) too, despite the fact that such systems are uncommon in nature. It has the form
\begin{eqnarray}
V_{Cou} = \sum_{i<j} \frac{C}{\mid \vec r_i - \vec r_j\mid};
\end{eqnarray}
where, $C$ is the strength of the interaction containing all the information about charge of particles and coefficients of the medium.
\section*{Wave function \& calculation procedures}
It is advantageous to investigate the bulk characteristics of FQHE in spherical geometry since it has no edges. Thus, we can define the composite fermion wave function in the typical spherical geometry  \cite{Haldane, Jain_book, Dirac} for our numerical computations. It is hypothesised that $N$ correlated atoms are present on a sphere's surface of radius $R$ under the influence of a radial magnetic field. A 'magnetic monopole' of strength $Q$ that is positioned at the centre of the sphere is assumed to be the source of the magnetic field. This monopole generates a total magnetic flux of $2Q\phi_0$ throughout the sphere's surface. So the radius of the sphere can be expressed as $R =\sqrt{Q} l_B$ ($l_B=\sqrt{\frac{\hbar c}{eB}}$ is magnetic length). This translates into a system of CFs with an effective flux $2q = 2Q - p(N - 1)$ \cite{Dev_Jain}. Here $Q$ has to be chosen to ensure that the state at $q$ is an integral quantum Hall state with an integer filling $n$; resulting in the same filling fraction $\nu$, same as equation (\ref{CF_seq}). The value of angular momentum ($L$) for a particle in the $n$-th LL is $n+Q-1$ \cite{monopole_harmonics}. For a single-particle the basis state is given by monopole harmonics \cite{Jain_kamilla}.
Thus the ground-state wave function at filling fraction $\nu$ for $N$ Bose atoms is given by \cite{wave_funcn},
\begin{equation}
  \Psi^0 = J^{-1} P_{LLL} J^2 \; \Phi_1(\Omega_1, \Omega_2, \cdots \Omega_N) \;.
\end{equation}
Here, the position of $i$-th particle on the spherical surface is $\Omega_i$, and $\Phi_1$ denotes the Slater determinant of fully filled $n$-number of $\Lambda$-levels of Bose-CFs. $P_{LLL}$ projects all the particles into the LLL \cite{PLL} and $J$ denotes the Jastrow factor \cite{DM_2014} having the form
\begin{eqnarray}
  J = \prod _{i<j}^N (u_i v_j - u_j v_i)^p  .\nonumber
\end{eqnarray}
We have used the spherical spinor-variables,
\begin{eqnarray*}
  u = cos(\theta/2) e^{-i\phi/2} \mbox{~~~ and ~~~~} v = sin (\theta/2) e^{i\phi/2} \;;
\end{eqnarray*}
where, $\theta$ and $\phi$ can vary from $0$ to $\pi$ and $0$ to $2\pi$ respectively. Following the CF transformation, the system's overall wave-function remains symmetric; since both $\Phi_1$ and $J$ are odd in this case. The ground state for filling fraction $\nu =1/2$ is shown in FIG. \ref{ground}, where we considered that the Zeeman energy ($E_z$) between the two spin-states is larger than the 3-times of cyclotron energy gap ($E_c$) then only we can expect collective excitations between same spin-component only (Other spin $\Lambda$-levels has no role here). At this condition the system behaves as a fully spin-polarized state. \\

A very significant physical quantity for an interacting system is it's pair-correlation function $g(r_{ij})$, which is basically the probability of having two particles at a separation distance $r_{ij}$ \cite{PC}. Let for a $N$-particle system with particle density $\rho$, the pair correlation function between particle $1$ and $2$ can be calculated as a function of relative coordinates between those two particles
{\small \begin{eqnarray}
  g(r_{12}) = \frac{N(N-1)}{\rho } \int d^{2}r_{3}d^{2}r_{4} \cdots d^{2}r_{N} \mid\Psi^0 (r_{1}, r_{2}, \cdots,r_{N})\mid^{2} \nonumber
\end{eqnarray} }

 In FQHE the many-body ground state wave function ($\Psi^0$) is normalized so that the pair-correlation function approaches to unity at a large separation distance. Generally, $g(r_{ij})$ stables to a constant value for large separation distance $r_{ij}$, which confirms that the wave-function describe a strongly interacting liquid-state. For long-range order crystalline state $g(r_{ij})$ oscillates throughout the large distance \cite{polarization_MI}. We have calculated the pair-correlation function for those three fractions $\nu = 1/2, 1/4, 1/6$ and plotted the result in FIG. \ref{PC}, which indicates that those three states are pure liquid-like.\\
\begin{figure*}
 \centering
  \includegraphics[width = 12.5cm]{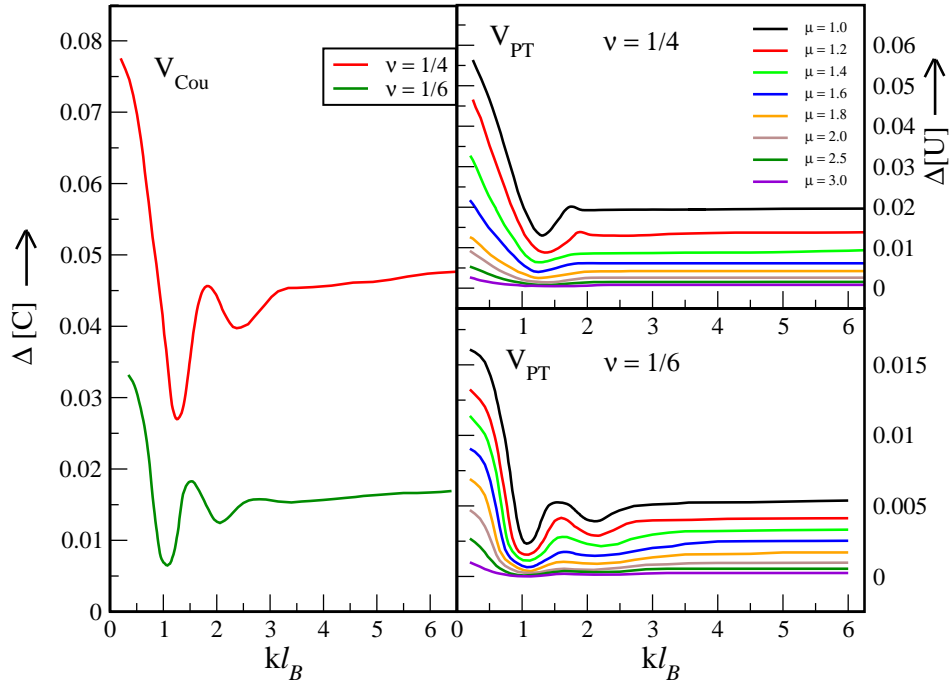}
  \caption{Spin-conserving excitation energy spectra for the filling fractions $\nu = 1/4, 1/6$. In the two right-most boxes, energy spectra (in unit of $U$) are shown using PT-interaction ($V_{PT}$) for different values of $\mu$. These $\mu$-values are increasing downwards (indicating range of interaction narrows down) in both of the filling fractions. Energy spectra (in unit of $C$) using long-range Coulomb interaction ($V_{cou}$) are depicted in the left-most box. Wave-vector ($k$) times $l_B$ is plotted along X-axis, which is connected to the total angular momentum as $k= L/R$.}
  \label{spectra}
 \end{figure*}

When a Bose-CF excites from the filled $0$-th $\Lambda$-level to an empty $n_f\;$-th $\Lambda$-level having same spin, then the excited state wave-function is given by \cite{DM_SSM} 
{\small{
\begin{equation}
  \Psi (L) = J^{-1} P_{LLL} J^2 \sum_{m_h} |m_h>\; <q, m_h; n_f+q,m_p|L,M> \;.
\end{equation}}}
Here $|m_h>$ represents the Slater determinant, where $N-1$ number of Bose-CF residing with a quasi-hole in the lowest $\Lambda$-level and one quasi-particle in the $n_f$-th $\Lambda$-level. The projection of angular momentum of the quasi-particle is $m_p$ and that of the quasi-hole is $m_h$.\; $<q, m_h;n_f+q,m_p|L,M>$ represent the well-known Clebsch-Gordan coefficients. The sum of all angular momenta is $L$. To avoid the numerical complexity, the Z-component of total angular momentum ($M=0$) is taken as zero, without losing generality.
 
 The superposition of all probable excitons yields the actual collective excitation rather than a single CF-exciton state. For spin-conserving excitations, three exciton states are taken into account 0 $\rightarrow$ 1;  0 $ \rightarrow  $ 2 and  0 $ \rightarrow  $ 3 i.e. one CF can experience this stimulation by hopping from the $0$-th $\Lambda$-level to a higher one. For $\nu = 1/2 $ filling fraction, the possible collective spin-conserving excitations are depicted in FIG. \ref{CDE}. Similar collective excitations are taken into account for the other two filling fractions. Nevertheless, for a given $L$, the wave functions for those three exciton states are not always orthogonal to one another. Since it is important to diagonalize the Hamiltonian in an orthonormal basis \cite{Mandal_Jain} in order to get significant results, we need to use a technique to orthogonalize the low-energy excited states. A well established technique known as Gram-Schmid orthonormalization procedure \cite{GS_ortho} can be used in the subspace of three states to derive the orthonormal basis. Then we have calculated the effective Hamiltonian matrix in the new orthogonal basis and diagonalize the effective Hamiltonian matrix to obtain the energies of the excited states.
The energy spectra i.e. the difference between the excited-state energy and the ground-state is obtained by 
\begin{equation}
  \Delta(L) = \frac{<\Psi(L)| H |\Psi(L)>}{<\Psi(L) |\Psi(L)>} - \frac{<\Psi^0| H |\Psi^0 >}{<\Psi^0 | \Psi^0 > }
  \label{Del_L}
\end{equation}
The first term of the R.H.S. of the above equation (\ref{Del_L}) is the excitation energy and the latter one is the ground state energy.
The Hamiltonian of the system in equation (\ref{H_boson}) will only be the interaction potential ($V$); as we have assumed that the particles are restricted into the LLL and their kinetic energies are quantized. Quantum Monte-Carlo approach was used for the numerical computation of the multidimensional integration. In a previous study, Moumita {\it et. al.} \cite{SDE_MI} demonstrated the variation of ground state energies for the three filling fractions using PT-interaction potential.
Here we calculated the excitation spectra ($\Delta$) for $\nu = 1/4, 1/6$ for both long-range and short-range interaction potential considering variation of range of interaction. The energy spectra are computed by Monte Carlo integration with large number of particles upto $N=160$ and then the fitted results are plotted in FIG. \ref{spectra}. Two roton-minima are found for both the two fractions in case of long-range interaction (leftbox in FIG. \ref{spectra}). The roton-minima for $\nu=1/6$  are obtained at lower wave-vector region ($1.08$ and $2.06$) and those are shifted to higher wave-vector region ($1.26$ and $2.38$) for $\nu = 1/4$. In case of short-range PT-interaction, we observed that when the value of $\mu$ is low, there is a long-range interaction, the roton-minima is highly sharp; however, when we raise the $\mu$-value, i.e. decrease the range of interaction, the roton-minima gets shallow and eventually disappears. Two roton-minima are prominent only for $\nu=1/6$ not for $\nu=1/4$ (right-most boxes in FIG. \ref{spectra}). But for electronic FQH-system, we observed only one roton-minima for the filling fraction $\nu=1/2p+1$ in both the lowest order and next higher order excitation mode \cite{JPTC}, that is not followed by bosonic FQH-system. From this fact, we can conclude that collective excitation in bosonic FQH-system strongly depends on the range of interaction between the particles.
\section*{Conclusion} 
Two roton-minima are obtained for both the two fractions in case of long-range Coulomb interaction. The roton-minima are obtained at the wave-vectors ($k l_B$) $1.26$ and $2.38$ for $\nu = 1/4$ and those are shifted to $1.08$ and $2.06$ for $\nu = 1/6$ in case of long-range interaction. But for short-range PT-interaction two roton-minima are more prominent for $\nu = 1/6$ only. Increase of flux attachment adds up the strong correlation between the particles to obtain two roton-minima even for short-range interaction in the case of $\nu = 1/6$. So the nature of the spectra in bosonic FQH-system, strongly depends upon the range of interaction as well as the number of attached flux quanta.  
\section*{Acknowledgement}
We thank the Department of Physics, IIEST Shibpur, Howrah, West Bengal for the computational facilities. 

\end{document}